\documentclass[12pt]{article}
\usepackage{epsfig}
\usepackage{graphicx}
\usepackage{amssymb}
\usepackage{amsmath}
\usepackage{amsfonts}
\usepackage{mathrsfs}
\usepackage[dvips]{color}

\newcommand{\R}{\mathbb{R}}

\newcommand{\fK}{\mathfrak{K}}

\newcommand{\cH}{\mathcal{H}}

\newcommand{\cP}{\mathcal{P}}
\newcommand{\cQ}{\mathcal{Q}}

\newcommand{\cT}{\mathcal{T}}

\newcommand{\be}{\begin{equation}}
\newcommand{\ee}{\end{equation}}
\newcommand{\bea}{\begin{eqnarray}}
\newcommand{\eea}{\end{eqnarray}}
\newcommand{\nn}{\nonumber}

\newcommand{\ed}{\end{document}}

\newcommand{\np}{\newpage}

\newcommand{\bi}{\begin{itemize}}
\newcommand{\ei}{\end{itemize}}

\newcommand{\bce}{\begin{center}}
\newcommand{\ece}{\end{center}}

\newcommand{\RE}{\,{\rm Re}}
\newcommand{\IM}{\,{\rm Im}}

\newcommand{\sH}{\mathscr{H}}
\newcommand{\sI}{\mathscr{I}}

\begin{document}

\title{A Hamiltonian Formulation of the Pais-Uhlenbeck Oscillator that
Yields a Stable and Unitary Quantum System}

\author{Ali~Mostafazadeh\thanks{E-mail address:
amostafazadeh@ku.edu.tr, Phone: +90 212 338 1462, Fax: +90 212 338
1559}
\\
Department of Mathematics, Ko\c{c} University, \\ 34450 Sar{\i}yer,
Istanbul, Turkey}

\date{ }
\maketitle

\begin{abstract}

We offer a new Hamiltonian formulation of the classical
Pais-Uhlenbeck Oscillator and consider its canonical quantization.
We show that for the non-degenerate case where the frequencies
differ, the quantum Hamiltonian operator is a Hermitian operator
with a positive spectrum, i.e., the quantum system is both stable
and unitary. A consistent description of the degenerate case based
on a Hamiltonian that is quadratic in momenta requires its analytic
continuation into a complex Hamiltonian system possessing a
generalized $\cP\cT$-symmetry (an involutive antilinear symmetry).
We devise a real description of this complex system, derive an
integral of motion for it, and explore its quantization.

\vspace{2mm}

\vspace{2mm}

\end{abstract}
\vspace{5mm}

\section{Introduction}
Pais-Uhlenbeck (PU) Oscillator \cite{PU} is the simplest and by far
the best known toy model for a higher derivative theory. The
interest in this model is motivated by the fact that it provides an
opportunity to explore the possibility of solving a long standing
problem related with the non-unitarity of higher derivative theories
(of gravity). These theories are particularly interesting because
they are known to be perturbatively renormalizable \cite{stelle}.

Consider the following equation of motion \cite{ma-da}.
    \be
    z^{(4)}+\alpha\,z^{(2)}+\beta\,z=0,
    \label{eq-mo}
    \ee
where $z$ is a real-valued function of time, $z^{(k)}$ denotes the
$k$-th derivative of $z$, and $\alpha$ and $\beta$ are real and
positive parameters related to a pair of frequencies $\omega_1$ and
$\omega_2$ according to
    \be
    \alpha:=\omega_1^2+\omega_2^2,~~~~~~\beta:=\omega_1^2\omega_2^2.
    \label{al-be}
    \ee
Throughout this article we define the classical PU oscillator as the
classical dynamical system determined by the forth-order equation of
motion (\ref{eq-mo}). Our aim is to obtain a unitary and stable
quantum system that has the classical system defined by
(\ref{eq-mo}) as its classical limit. Here by unitarity and
stability of a quantum system we mean that the corresponding quantum
Hamiltonian operator is Hermitian and its spectrum is bounded from
below.

It is well-known that (\ref{eq-mo}) can be derived from a Lagrangian
of the form \cite{PU,ma-da}
    \be
    L=\frac{\mu}{2}(\ddot z^2-\alpha\dot z^2+\beta z^2),
    \label{lag}
    \ee
where $\mu$ is an arbitrary positive real mass parameter and each
over-dot stands for a time-derivative. Introducing a new coordinate
variable $x$ according to $x=\dot z$, one can turn (\ref{lag}) into
the Lagrangian of a constraint second derivative theory and apply
the machinery of Dirac's constraint quantization to arrive at a
Hamiltonian description of the PU Oscillator based on the following
classical Hamiltonian \cite{ma-da,smilga-plb}.
    \be
    H=\frac{p_x^2}{2\mu}+xp_z+\frac{\mu\alpha x^2}{2}-
    \frac{\mu\beta z^2}{2}.
    \label{H1}
    \ee

In Ref.~\cite{be-ma}, the authors describe how the standard
quantization of the classical PU oscillator based on the Hamiltonian
(\ref{H1}) gives rise to a theory that is either non-unitary or
unstable. They then take the difficult-to-justify step of requiring
that the coordinate $z$ be considered as imaginary while the
coordinate $x$, that is related to $z$ via $x=\dot z$, is treated as
being real. This corresponds to a drastic change of boundary
conditions that define the Hilbert space of the quantum theory.
Using a similarity transformation they rotate the complex $z$-plane
($z\to y:=iz$) to map the Hilbert space defined by this unusual
boundary conditions to the usual Hilbert space $L^2(\R^2)$. However,
in this Hilbert space the Hamiltonian takes the form
    \be
    H=\frac{p_x^2}{2\mu }+\frac{\mu\alpha x^2}{2}+
    \frac{\mu\beta y^2}{2}-ip_yx,
    \label{H2}
    \ee
which is manifestly non-Hermitian and $\cP\cT$-symmetric (the action
of $\cP$ being defined by $x\to -x$, $p_x\to -p_x$, $y\to y$, and
$p_y\to p_y$). It turns out that (\ref{H2}) is quasi-Hermitian
\cite{quasi}, so that it can be mapped using another similarity
transformation to a Hermitian Hamiltonian \cite{jpa-2003}. Using
such a similarity transformation, one finds the Hamiltonian for an
uncoupled pair of harmonic oscillators, namely
$H=\left(\frac{p_x^2}{2\mu }+\frac{\mu\omega_1^2 x^2}{2}\right)+
    \left(\frac{p_y^2}{2\mu\omega_1}+\frac{\mu\beta y^2}{2}\right)$,
that is clearly Hermitian and has a positive spectrum. The authors
of \cite{be-ma} consider this procedure as a proof of a ``No-Ghost
Theorem for the Forth-Order Derivative PU Oscillator Model,'' though
they admit that the quantum system they obtain does not have the
classical PU oscillator (\ref{eq-mo}) as its classical limit!

The basic idea of the present investigation is that the forth-order
equation of motion (\ref{eq-mo}) that we use to define the classical
PU oscillator may be derived using other Hamiltonians  in the
Hamiltonian formulation of classical mechanics. This suggests
searching for a new classical Hamiltonian that generates the
dynamical equation (\ref{eq-mo}) and achieves the same goals as
those of the approach of Ref.~\cite{be-ma}, but avoids treating $z$
as an imaginary variable. We will construct such a classical
Hamiltonian for the non-degenerate PU Oscillator (where
$\omega_1\neq\omega_2$) and elaborate on the degenerate case.

\section{An Alternative Hamiltonian Formulation}

The starting point of our analysis is the introduction of a new
variable, namely
    \be
    w:=\tau^2(\ddot z+\lambda z),
    \label{w=}
    \ee
where $\tau$ and $\lambda$ are a pair of nonzero free real
parameters with dimension of time$^2$ and time$^{-2}$, respectively,
so that $w$ and $z$ have the same dimension. It is an easy exercise
to show that (\ref{eq-mo}) is equivalent to
    \bea
    && \ddot z=\tau^{-2}w-\lambda z,
    \label{e1}\\
    && \ddot w=
    (\lambda-\alpha)w-\tau^2(\lambda^2-\alpha\lambda+\beta)z.
    \label{e2}
    \eea
We can view these equations as Newton's equations of motion for a
system with two degrees of freedom.

Next, we wish to obtain a Hamiltonian formulation of the dynamical
system defined by (\ref{e1}) and (\ref{e2}). As a first step we
examine conditions on the free parameters $\tau$ and $\lambda$ for
which the system is subject to a conservative force. Introducing
mass parameters $\mu_z$, $\mu_w$, and demanding the existence of a
potential field $V$ such that (\ref{e1}) and (\ref{e2}) take the
form $\mu_w\ddot w=-\partial_w V$ and $\mu_z\ddot z=-\partial_z V$,
we find by imposing the integrability condition,
$\partial_w\partial_z V=\partial_z\partial_w V$, that
    \be
    \lambda^2-\alpha\lambda+\beta+\frac{\mu_z}{\mu_w\tau^4}=0.
    \label{e3}
    \ee
Introducing
    \be
    \Omega:=(4\mu_w^{-1}\mu_z)^{1/4}\tau^{-1},~~~~
    \delta:=\alpha^2-4\beta-\Omega^4=(\omega_1^2-\omega_2^2)^2-\Omega^4,
    \label{om-de}
    \ee
we can express the solutions of (\ref{e3}) in the form
    \be
    \lambda=\frac{\alpha\pm\sqrt\delta}{2}.
    \label{lambda=}
    \ee
With these choices for $\lambda$, we can determine the potential
field $V$, and introducing the momentum variables $p_w:=\mu_w\dot w$
and $p_z:=\mu_z\dot z$, we obtain the following one-parameter family
of classical Hamiltonians
    \be
    H_\delta=\frac{p_w^2}{2\mu_w}+\frac{p_z^2}{2\mu_z}+
    \frac{\mu_w}{4}(\alpha\mp\sqrt\delta)w^2+
    \frac{\mu_z}{4}(\alpha\pm\sqrt\delta)z^2-
    \frac{1}{2}\sqrt{\mu_w\mu_z(\alpha^2-4\beta-\delta)}\,wz.
    \label{H4}
    \ee
Note that because we have not yet fixed the value of $\tau$ and
consequently $\Omega$, the parameter $\delta$ is a free parameter.
For all values of $\delta$, the Hamilton equations associated with
$H_\delta$ are equivalent to the the equation of motion
(\ref{eq-mo}) that defines the classical PU oscillator. A natural
question is whether there are values of $\delta$ for which the
quantization of $H_\delta$ would yield a stable and unitary quantum
system. This turns out to require a separate analysis for the
degenerate ($\omega_1=\omega_2$) and non-degenerate
($\omega_1\neq\omega_2$) cases.

\section{Non-degenerate PU Oscillator and Its Quantization}

As seen from (\ref{H4}), $H_\delta$ is a real-valued function on the
phase space provided that $\delta\geq 0$ that is
$\sqrt{|\omega_1^2-\omega_2^2|}\geq\Omega$. This is only possible
for the non-degenerate case where $\omega_1\neq\omega_2$. We will
next show that in this case $H_\delta$ is bounded from below.

First we note that according to (\ref{al-be}) and (\ref{om-de}),
$\alpha\pm\sqrt\delta>0$. This allows us to introduce the real and
positive parameters:
    \be\nu_w:=\frac{1}{2}\sqrt{\mu_w(\alpha\mp\sqrt\delta)},~~~~
        \nu_z:=\frac{1}{2}\sqrt{\mu_z(\alpha\pm\sqrt\delta)},
        \ee
and express (\ref{H4}) as
    \bea
     H_\delta&=&\frac{p_w^2}{2\mu_w}+\frac{p_z^2}{2\mu_z}+
    \nu_w^2w^2+
    \nu_z^2z^2-\frac{2\nu_w\nu_z\Omega^2wz}{\sqrt{\alpha^2-\delta}}\nn\\
    &=&\frac{p_w^2}{2\mu_w}+\frac{p_z^2}{2\mu_z}+
    \left(\nu_ww-\frac{\nu_z\Omega^2z}{\sqrt{\alpha^2-\delta}}\right)^2+
    \frac{4\beta \nu_z^2z^2}{\alpha^2-\delta}.
    \label{H5}
    \eea
In view of (\ref{al-be}) and (\ref{om-de}), the last term on the
right-hand side of (\ref{H5}) is nonnegative. Therefore, $H_\delta$
is bounded below by zero for all possible values of $\omega_1$ and
$\omega_2\neq\omega_1$ (where $\delta\geq 0$).

We can easily quantize the Hamiltonian (\ref{H5}) by applying the
standard canonical quantization scheme:
    \be
    (w,z,p_w,p_z)\to(\hat w,\hat z,\hat p_w,\hat p_z),
    \label{q0}
    \ee
where $\hat w,\hat z,\hat p_w$, and $\hat p_z$ are operators acting
in $L^2(\R^2)$ as follows. For all $\psi\in L^2(\R^2)$,
    \bea
    \hat w\psi(w,z)=w\psi(w,z),~~~~~~
    \hat z\psi(w,z)=z\psi(w,z),~~~~~~~~~~~~~~
    \label{q1}\\
    \hat p_w\psi(w,z)=-i\hbar\partial_w\psi(w,z),~~
    \hat p_z\psi(w,z)=-i\hbar\partial_z\psi(w,z).~~~
    \label{q2}
    \eea
The resulting quantum Hamiltonian operator,
    \be
    \hat{H_\delta}=\frac{\hat p_w^2}{2\mu_w}+\frac{\hat p_z^2}{2\mu_z}+
    \left(\nu_w\hat w-\frac{\nu_z\Omega^2\hat z}{\sqrt{\alpha^2-\delta}}
    \right)^2+\frac{4\beta \nu_z^2\hat z^2}{\alpha^2-\delta},
    \label{H-z10}
    \ee
that also acts in the Hilbert space $L^2(\R^2)$, is manifestly
Hermitian, and being the sum of positive operators, it is a positive
operator (with a non-negative spectrum.) It describes a stable and
unitary quantum system consisting of a coupled pair of harmonic
oscillators. By construction, taking the classical limit of this
quantum system we recover the classical PU oscillator as defined by
the equation of motion (\ref{eq-mo}).

Because (\ref{H-z10}) is a quadratic Hamiltonian, we may try to
decouple and diagonalize it using a linear canonical transformation.
We defer the details to the appendix and suffice to mention that,
for the non-degenerate PU oscillator that we consider here, there is
a similarity transformation generated by a quadratic function $\hat
Q$ of $\hat w,\hat z,\hat p_w$, and $\hat p_z$ that maps
$\hat{H_\delta}$ to the Hamiltonian operator $\hat{H'_\delta}$ for a
pair of decoupled harmonic oscillators:
    \be
    \hat{H_\delta}\to \hat{H'_\delta}=e^{-\hat Q/2}\hat{H_\delta}\,
    e^{\hat Q/2}=\frac{\hat p_w^2}{2\mu_w'}+
    \frac{\mu_w'\omega_-^2}{2}\,\hat w^2+
    \frac{\hat p_z^2}{2\mu_z'}+
    \frac{\mu_z'\omega_+^2}{2}\,\hat z^2,
    \ee
where $\mu_w',\mu_z',\omega_\pm$ are positive real parameters whose
explicit form are given in (\ref{new-para}), below.

\section{Degenerate PU Oscillator and Its Quantization}

For the degenerate PU oscillator, where
    \be
    \omega_1=\omega_2=:\omega,~~~~\alpha=2\omega^2,~~~~
    \delta=-\Omega^4,~~~~\lambda=\omega^2\pm\frac{i\Omega^2}{2},
    ~~~~\tau=\frac{\sqrt 2}{\Omega},
     \label{deg}
    \ee
the Hamiltonian (\ref{H4}) reads
    \bea
    H_{-\Omega^4}&=&\frac{p_w^2}{2\mu_w}+\frac{p_z^2}{2\mu_z}+
    \frac{\mu_w}{4}(2\omega^2\mp i\Omega^2)w^2+\frac{\mu_z}{4}(2\omega^2\pm
    i\Omega^2)z^2-\frac{1}{2}\sqrt{\mu_w\mu_z}\,\Omega^2\, w z.
    \label{H6}
    \eea
If we choose the mass parameters, that are actually not constrained
by the classical equation of motion~(\ref{eq-mo}), to coincide,
i.e., set $\mu_w=\mu_z=:\mu$, the Hamiltonian (\ref{H6}) takes the
form
    \bea
    H_{-\Omega^4}&=&\frac{1}{2\mu}(p_w^2+p_z^2)+
    \frac{\mu}{4}\left[(2\omega^2\mp i\Omega^2)w^2+(2\omega^2\pm
    i\Omega^2)z^2\right]-\frac{\mu\,\Omega^2 w z}{2}.
    \label{H7}
    \eea
Note that here $\Omega$ is a free nonzero real parameter.

It is easy to see that regardless of the value of $\Omega$, the
Hamiltonian $H_{-\Omega^4}$ is invariant under the combined effect
of complex-conjugation and swapping $(w,p_w)$ and $(z,p_z)$.
Quantizing (\ref{H7}) using the standard canonical quantization
scheme (\ref{q0}), we obtain a non-Hermitian Hamiltonian operator
(acting in $L^2(\R^2)$), namely
    \bea
    \hat H_{-\Omega^4}&=&\frac{1}{2\mu}(\hat p_w^2+\hat p_z^2)+
    \frac{\mu}{4}\left[(2\omega^2\mp i\Omega^2)\hat w^2+(2\omega^2\pm
    i\Omega^2)\hat z^2\right]-\frac{\mu\,\Omega^2 \hat w \hat z}{2}.
    \label{H7q}
    \eea
that possesses a particular antilinear symmetry. This is a
generalized $\cP\cT$-symmetry whose generator we denote by $\cQ\cT$,
\cite{jpa-2008b}. Here $\cQ$ is the operator of reflection about the
line $w=z$ in the $w$-$z$ plane, and $\cT$ is the time-reversal
operator; for all $\psi\in L^2(\R^2)$, $(\cQ\psi)(w,z)=\psi(z,w)$
and $(\cT\psi)(w,z)=\psi(w,z)^*$.

A major difference between our Hamiltonian formulation of the
non-degenerate and degenerate PU oscillators is that in the latter
case the classical Hamiltonian (\ref{H6}) is a complex-valued
function of the phase space variables $(w,z,p_w,p_z)$. Because
$\Omega$ is not constrained, we can take $\Omega\ll \omega$ and try
to use perturbation theory to see if the the spectrum of the
Hamiltonian $\hat H_{-\Omega^4}$ is real, \cite{cal}. Again, the
fact that $\hat H_{-\Omega^4}$ is a quadratic this Hamiltonian
suggests the possibility of its diagonalization via a (possibly
complex) linear canonical transformation. As we show in the appendix
this turns out not to be possible. Indeed the lack of such a
canonical transformation might be an indication that, similarly to
the coupled oscillators studied in \cite{CGS}, $\hat H_{-\Omega^4}$
is actually non-diagonalizable. We shall not pursue the study of the
spectral properties of this Hamiltonian operator here, because as we
explain below it does not define a quantum system that has the
classical degenerate PU oscillator as its classical limit.

Note that in view of (\ref{deg}) and (\ref{al-be}), the parameter
$\lambda$ and the coordinate variable $w$ actually take complex
values. This observation, which is often missed or ignored in the
study of the classical systems underlying $\cP\cT$-symmetric quantum
systems \cite{bender-class}, reveals a basic deficiency of a
quantization scheme involving $w\to\hat w$, simply because while $w$
is a complex variable the operator $\hat w$, that is defined in
(\ref{q1}), has a real spectrum. Performing the quantization of
(\ref{H6}) using (\ref{q0}) does yield a non-Hermitian Hamiltonian
operator $\hat H_{-\Omega^4}$ with an antilinear symmetry, but even
if this Hamiltonian turns out to be quasi-Hermitian and capable of
defining a unitary quantum system, this system does not admit the
classical PU oscillator as its classical limit. The same conclusion
applies to the analysis of \cite{be-ma}. The main difference is that
in our approach this problem only arises for the degenerate PU
oscillator, while in the approach of Ref.~\cite{be-ma} it is present
also for the non-degenerate PU oscillator.

Next, we recall that in view of (\ref{deg}), Eqs.~(\ref{eq-mo}),
(\ref{e1}), and (\ref{e2}) take the form
    \bea
    &&z^{(4)}+2\omega^2 z^{(2)}+\omega^4z=0,
    \label{eq-mo-deg}\\
    &&\mu\ddot z=kw-(k\pm i\fK)z,
    \label{e1d}\\
    &&\mu\ddot w=(-k\pm i\fK)w+k z,
    \label{e2d}
    \eea
where $k:=\mu\,\omega^2$ and $\fK:=\frac{\mu\,\Omega^2}{2}$. Note
that, as in the non-degenerate case, here $z$ is a real-valued
function of time, whereas $w$ is necessarily complex-valued. This is
because (\ref{e1d}) and (\ref{e2d}) yield (\ref{eq-mo-deg}) only for
$\fK\neq 0$.

We can express Eqs.~(\ref{e1d}) and (\ref{e2d}) in terms of real and
imaginary parts of $w$, namely $w_1:=\RE(w)$ and $w_2:=\IM(w)$.
These yield a system of three Newton's equations for $z,w_1$ and
$w_2$ that can be reduced to
    \be
    \mu\,\ddot z=\fK w_1-kz,~~~\mu\,\ddot w_1=-kw_1,~~~w_2=\pm z.
    \ee
It is not difficult to show that the force corresponding to the
first two of these equations is nonconservative. This means that
this system does not admit a Hamiltonian formulation based on a
Hamiltonian that is quadratic in momenta. In other words, we can
obtain a consistent Hamiltonian formulation of the classical
dynamics of the system using the Hamiltonian (\ref{H7}), only if we
treat both the coordinate variables $w$ and $z$ and the
corresponding momentum variables $p_w$ and $p_z$ as complex
variables. This yields an analytic continuation of the degenerate PU
oscillator into a complex dynamical system defined by (\ref{H7}) via
the complex Hamilton equations
    \be
    \dot w =\partial_{p_w}H,~~~~
    \dot z =\partial_{p_z}H,~~~~
    \dot p_z =-\partial_{z}H,~~~~
    \dot p_w =-\partial_{w}H,
    \label{ha-eq}
    \ee
where we use $H$ in place of $H_{-\Omega^4}$ for simplicity.

A detailed study of complex classical systems of the type
(\ref{ha-eq}) (though with a single complex coordinate variable) has
been carried out in \cite{pla-2006}. Equations (\ref{ha-eq}) are
meaningful if $H$ is a complex analytic function of $w$, $z$, $p_w$,
and $p_z$. This implies that Cauchy-Riemann conditions hold for the
real and imaginary parts of $H$. That is, introducing
    \bea
    &&z_1:=\RE(z),~~~~z_2:=\IM(z),~~~~
    p_{z_1}:=\RE(p_z),\nn\\
    &&p_{z_2}:=\IM(p_z),~~~~
    p_{w_1}:=\RE(w_z),~~~~p_{w_2}:=\IM(w_z),\nn\\
    && H_1:=\RE(H),~~~~H_2:=\IM(H),
    \nn
    \eea
we have
    \bea
    &&\partial_{w_1}H_1=\partial_{w_2}H_2,~~~~
    \partial_{w_2}H_1=-\partial_{w_1}H_2,
    \label{CR-1}\\
    &&\partial_{p_{w_1}}H_1=\partial_{p_{w_2}}H_2,~~~~
    \partial_{p_{w_2}}H_1=-\partial_{p_{w_1}}H_2,
    \label{CR-12}\\
    &&\partial_{z_1}H_1=\partial_{z_2}H_2,~~~~
    \partial_{z_2}H_1=-\partial_{z_1}H_2,
    \label{CR-13}\\
    &&\partial_{p_{z_1}}H_1=\partial_{p_{z_2}}H_2,~~~~
    \partial_{p_{z_2}}H_1=-\partial_{p_{z_1}}H_2.
    \label{CR-2}
    \eea

As originally noted in \cite{pla-2006} for one-dimensional complex
configuration spaces, the following are miraculous consequences of
the Cauchy-Riemann conditions (\ref{CR-1}) -- (\ref{CR-2}).
    \begin{itemize}
    \item Eqs.~(\ref{ha-eq}) are equivalent to a system of
    Hamilton's equations, $
        \dot x_j=\partial_{p_j}\sH$, $\dot p_j=-\partial_{x_j}\sH$,
    for the phase space variables
        \be
        \begin{array}{c}
        x_1:=w_1,~~~~x_2=p_{w_2},~~~~x_3:=z_1,~~~~x_4:=p_{z_2},\\
        p_1:=p_{w_1},~~~~p_2=w_2,~~~~p_3:=p_{z_1},~~~~p_4:=z_2,
        \end{array}\nn
        \ee
    where $\sH$ is nothing but the real part of the complex
    Hamiltonian $H$, i.e., $\sH:=H_1$.
    \item The imaginary part $\sI:=H_2$ of the complex Hamiltonian $H$ is an
    integral of motion for the system that is independent of $\sH$,
        \[\{\sI,\sH\}:=\sum_{j=1}^4\left[(\partial_{x_j}\sI)(\partial_{p_j}\sH)
    -(\partial_{p_j}\sI)(\partial_{xj}\sH)\right]=0.\]

    \end{itemize}

The problem of quantizing one-dimensional examples of similar
complex dynamical systems has been considered in \cite{smilga}. The
integral of motion $\sI$ generates a type of gauge symmetry for
these systems \cite{pla-2006,smilga} that makes their quantization a
nontrivial and non-unique process. Here we will only derive the
explicit form of $\sH$ and $\sI$ to examine if a canonical
quantization of $\sH$ (before imposing the constraint) can give rise
to a stable and unitary dynamics. A straightforward calculation
gives
    \bea
    \sH&=&\frac{p_1^2+p_3^2}{2\mu}-\frac{k}{2}(p_2^2+p_4^2)+
    \fK p_2p_4\pm\fK(x_1p_2-x_3p_4)-\frac{x_2^2+x_4^2}{2\mu}+
    \frac{k}{2}(x_1^2+x_3^2)-\fK\, x_1x_3,~~~~~~
    \label{H1=}
    \\
    \sI&=&\pm\frac{\fK}{2}(p_2^2-p_4^2)+\frac{x_2p_1+x_4p_3}{2\mu}+
    k(x_1p_2+x_3p_4)-\fK(x_1p_4+x_3p_2)\pm\frac{\fK}{2}(x_3^2-x_1^2).
    \label{H2=}
    \eea
As seen from (\ref{H1=}), $\sH$ is an unbounded function of the
phase space variables. Therefore, the quantization of $\sH$ leads to
a Hermitian operator $\hat\sH$ whose spectrum is not bounded below.
However, note that to impose the constraint $\Phi:=\sI-C=0$, where
$C$ is a real constant, we need to quantize $\Phi$ to obtain the
quantum constraint $\hat\Phi$ and define the physical Hilbert space
of the system as the null space (kernel) of $\hat\Phi$. Given the
complicated form of $\hat\sH$ and $\hat\Phi$, it is not easy to
determine if the dynamics taking place in the physical Hilbert space
is stable. But this seems to be unlikely.

Alternatively we can impose the constraint before quantization by
solving $\Phi=0$ for one of the variables and substituting the
result in the expression for $\sH$. This yields a highly complicated
reduced classical Hamiltonian that is still unbounded both from
above and below. Hence its quantization gives rise to an unstable
quantum system.

\section{Concluding Remarks}

We have established a novel Hamiltonian formulation of the
Pais-Uhlenbeck oscillator that yields a real and positive classical
Hamiltonian for a non-degenerate oscillator. Canonical quantization
of this Hamiltonian yields a unitary and stable quantum system with
the classical PU oscillator as its classical limit. This provides a
simple and consistent solution of the long-standing non-unitarity
versus instability problem that one encounters in quantizing the
classical PU Oscillator.

Our Hamiltonian formulation involves a manifestly complex classical
Hamiltonian whenever the PU oscillator is degenerate. The naive
canonical quantization of this Hamiltonian, that allows for mapping
complex classical variables to Hermitian operators, gives rise to a
non-Hermitian Hamiltonian operator that possesses a generalized
$\cP\cT$-symmetry. We are of the opinion that even through this
Hamiltonian has appealing symmetry properties, it is not relevant to
the problem of quantizing the classical PU oscillator. This is
because this approach is void of a consistent quantum-to-classical
correspondence that would map classical observables to quantum
observables.

A consistent method of dealing with this complex classical
Hamiltonian is to realize that it admits a real description. Using
this description one discovers that the dynamics is actually
generated by the real part $\sH$ of the complex Hamiltonian $H$ and
that the system has a highly nontrivial integral of motion given by
the imaginary part $\sI$ of $H$. The latter may be identified both
as the generator of a gauge symmetry and as a first class
constraint. A proper quantization of this system must take into
account the presence of this constraint. The Hamiltonian operator
$\hat\sH$ that one obtains by quantizing $\sH$ is a Hermitian
operator with an unbounded spectrum both from above and below. In
principle this is not sufficient to conclude that the quantum
dynamics is unstable, because the physical Hilbert space $\cH_{\rm
phys}$ is the null space of the quantum constraint $\hat\Phi=\hat
\sI-C$. To determine the stability of the quantum system after the
imposition of the constraint, one must study the spectrum of the
restriction of $\hat\sH$ to $\cH_{\rm phys}$. Given the complicated
form of $\hat\sH$ and $\hat\Phi$ this is not an easy task. There are
however hints that the degenerate quantum PU oscillator probably
remains unstable even after enforcing the constraint. The presence
of runaway classical solutions seems to support this
conclusion.\vspace{.3cm}

\noindent {{\textbf{Acknowledgment:}}} This work has been supported
by the Turkish Academy of Sciences (T\"UBA).

\section*{Appendix}

In this appendix we explore the possibility of decoupling and
diagonalizing the Hamiltonian operator obtained by standard
canonical quantization (\ref{q0}) of (\ref{H4}), namely
    \be
    \hat H_\delta=\frac{\hat p_w^2}{2\mu_w}+\frac{\hat p_z^2}{2\mu_z}+
    \frac{\mu_w}{4}(\alpha\mp\sqrt\delta)\hat w^2+
    \frac{\mu_z}{4}(\alpha\pm\sqrt\delta)\hat z^2-
    \frac{1}{2}\sqrt{\mu_w\mu_z(\alpha^2-4\beta-\delta)}\,\hat w\hat z.
    \label{H4q}
    \ee
First, we introduce the dimensionless quantities:
     \bea
     &&\hat x_1:=\sqrt{\frac{\mu_w\sqrt\alpha}{\hbar}}\:\hat w,~~~
     \hat x_2:=\sqrt{\frac{\mu_z\sqrt\alpha}{\hbar}}\hat z,~~~
     \hat p_1:=\frac{\hat p_w}{\sqrt{\hbar\mu_w\sqrt\alpha}},~~~
     \hat p_2:=\frac{\hat p_z}{\sqrt{\hbar\mu_z\sqrt\alpha}},~~~
     \label{param1}\\
     &&\epsilon:=\pm\frac{\sqrt{\delta}}{\alpha}=
     \pm\frac{\sqrt{(\omega_1^2-\omega_2^2)^2-\Omega^4}}{
     \omega_1^2+\omega_2^2},~~~~~
     \gamma:=\left|1-\frac{4\beta}{\alpha^2}\right|^{\frac{1}{2}}=\left|
     \frac{\omega_1^2-\omega_2^2}{\omega_1^2+\omega_2^2}\right|,
     \label{param2}
     \eea
that allow us to express $\hat H_\delta$ in the form
    \be
    \hat H_\delta=\frac{\hbar\sqrt\alpha}{2}\left[\hat p_1^2+\hat
    p_2^2+\frac{1}{2}(1-\epsilon)\,\hat x_1^2+
    \frac{1}{2}(1+\epsilon)\,\hat x_1^2-
    \sqrt{\gamma^2-\epsilon^2}\,\hat x_1 \hat x_2\right].
    \label{H4n}
    \ee
It is easy to check that $[\hat x_i,\hat p_j]=i\delta_{ij}$ for
$i,j=1,2$, and $\sqrt{\gamma^2-\epsilon^2}=\Omega^2/\alpha\geq 0$.

A linear canonical transformation is equivalent to the similarity
transformation $\hat H_\delta\to \hat H':=e^{-\hat Q/2}\hat H_\delta
e^{\hat Q/2}$ where $\hat Q$ is a quadratic function of the
operators $\hat x_i$ and $\hat p_i$ with $i=1,2$. It is well-known
that the transformed operators $\hat X_i:=e^{-\hat Q/2}\hat x_i
e^{\hat Q/2}$ and $\hat P_i:=e^{-\hat Q/2}\hat p_i e^{\hat Q/2}$ are
linear functions of $\hat x_i$ and $\hat p_i$ satisfying the
canonical commutation relations
    \be
    [\hat X_i,\hat P_j]=i\delta_{ij}.
     \label{ccr}
    \ee

Let us introduce
    $\mathbf{\hat q}:=(\hat p_1,\hat x_1,\hat p_2,\hat x_2)^T$ and
    $\mathbf{\hat q'}:=(\hat P_1,\hat X_1,\hat P_2,\hat X_2)^T$,
where the superscript ``$T$'' stands for the transpose. Then we can
express $\hat H_\delta$ and $\hat H'_\delta$ in the form
    \be
    \hat H_\delta= \mathbf{\hat q}^T\mathbf{H}_\delta\,\mathbf{\hat
    q},~~~~~~~
    \hat H'_\delta= \mathbf{\hat{q'}}^{T}\mathbf{H}_\delta\,\mathbf{\hat q}',
    \label{HH}
    \ee
where
    \be
    \mathbf{H}_\delta:=\frac{\hbar\sqrt\alpha}{2}\left(\begin{array}{cccc}
    1 & 0 & 0 & 0\\
    0 & \frac{1}{2}(1-\epsilon) & 0 & -\frac{1}{2}\sqrt{\gamma^2-\epsilon^2}\\
    0 & 0 & 1 & 0\\
    0 & -\frac{1}{2}\sqrt{\gamma^2-\epsilon^2} & 0 &
    \frac{1}{2}(1+\epsilon)\end{array}\right).
    \label{bold-H}
    \ee
The linearity of the canonical transformation is equivalent to the
existence of an invertible $4\times 4$ complex matrix $\mathbf{U}$
(with numerical entries) such that
        \be
        \mathbf{\hat q'}=\mathbf{U}\mathbf{\hat q}.
        \label{LCT-1}
        \ee
In terms of $\mathbf{U}$ the canonical commutation relations
(\ref{ccr}) take the form
    \be
    \mathbf{U}^T\mathbf{C}\mathbf{U}=\mathbf{C},
    \label{ccr2}
    \ee
where
    \be
    \mathbf{C}:=\left(\begin{array}{cccc}
    0 & -1 & 0 & 0\\
    1 & 0 & 0 & 0\\
    0 & 0 & 0 & -1\\
    0 & 0 & 1 & 0\end{array}\right).
    \label{bold-H}
    \ee

We wish to find out if we can choose $\hat Q$ such that $\hat
H'_\delta$ is the Hamiltonian operator for a pair of decoupled
oscillators. In view of (\ref{HH}), this is equivalent to looking
for a matrix $\mathbf{U}$ such that the matrix
    \be
    \mathbf{H}'_\delta:=\mathbf{U}^T\mathbf{H}_\delta\mathbf{U}
    \label{H-prime}
    \ee
is diagonal. This problem has been studied in \cite{BCF}. Here we
outline its solution.

Using (\ref{H-prime}) and the properties of $\mathbf{C}$, we find
that
    \be
    \mathbf{(CH_\delta)^2U=U(-CH'_\delta C^TH'_\delta)}.
    \label{id}
    \ee
Now, for a diagonal $\mathbf{H'_\delta}$, the matrix
$-\mathbf{CH_\delta'C^TH_\delta'}$ is also diagonal, and (\ref{id})
implies that the columns of $\mathbf{U}$ are eigenvectors of
$(\mathbf{CH_\delta})^2$. We can easily compute and solve the
eigenvalue problem for this matrix and construct $\mathbf{U}$. The
result is
    \bea
    \mathbf{U}&=&\left(\begin{array}{cccc}
    0 & \frac{\epsilon+\gamma}{\sqrt{\gamma^2-\epsilon^2}}&
    0&\frac{\epsilon-\gamma}{\sqrt{\gamma^2-\epsilon^2}}\\
    \frac{\epsilon+\gamma}{\sqrt{\gamma^2-\epsilon^2}}& 0 &
    \frac{\epsilon-\gamma}{\sqrt{\gamma^2-\epsilon^2}}& 0\\
    0 & 1 & 0 &1\\
    1& 0 & 1 &0
    \end{array}\right),
    \label{U=}
    \eea

Substituting (\ref{U=}) in (\ref{H-prime}), we have checked that
indeed $\mathbf{H}'_\delta$ is a diagonal matrix with diagonal
entries:
    \be
    (\mathbf{H}'_\delta)_{11}=
    \frac{\hbar\sqrt\alpha\,\gamma(1-\gamma)}{2(\gamma-\epsilon)},~~~~
    (\mathbf{H}'_\delta)_{22}=
    \frac{\hbar\sqrt\alpha\,\gamma}{\gamma-\epsilon},~~~~
    (\mathbf{H}'_\delta)_{33}=
    \frac{\hbar\sqrt\alpha\,\gamma(1+\gamma)}{2(\gamma+\epsilon)},~~~~
    (\mathbf{H}'_\delta)_{44}=
    \frac{\hbar\sqrt\alpha\,\gamma}{\gamma+\epsilon}.
    \label{H44=}
    \ee
This implies that $\hat H'_\delta$ describes the dynamics of a pair
of decoupled oscillators:
    \bea
    \hat H'_\delta&=&{\mathbf{q}'}^T\mathbf{H}_\delta \mathbf{q}'=
    {\mathbf{q}}^T\mathbf{H}'_\delta\mathbf{q}=
    \frac{\hbar\sqrt\alpha}{2}\left[
    \frac{\gamma(1-\gamma)\hat p_1^2}{\gamma-\epsilon}+
    \frac{2\gamma\,\hat x_1^2}{\gamma-\epsilon}+
    \frac{\gamma(1+\gamma)\hat p_2^2}{\gamma+\epsilon}+
    \frac{2\gamma\,\hat x_2^2}{\gamma+\epsilon}\right],\nn\\
    &=&\frac{\hat p_w^2}{2\mu_w'}+
    \frac{\mu_w'\omega_-^2}{2}\,\hat w^2+
    \frac{\hat p_z^2}{2\mu_z'}+
    \frac{\mu_z'\omega_+^2}{2}\,\hat z^2,
    \label{trans-H}
    \eea
where
    \be
    \mu_w':=\frac{(\gamma-\epsilon)\mu_w}{\gamma(1-\gamma)},~~~
    \mu_z':=\frac{(\gamma+\epsilon)\mu_z}{\gamma(1+\gamma)},~~~
    \omega_\pm:=\frac{\sqrt{2\gamma^2(1\pm\gamma)\alpha}}{\gamma\pm\epsilon}.
    \label{new-para}
    \ee
For the non-degenerate PU oscillator where $\delta$ is a nonnegative
real number, $\mu'_w,\mu'_z$ and $\omega_\pm$ are real and positive
parameters, and $\hat H'_\delta$ is the sum of two noninteracting
simple harmonic oscillators. Clearly the spectrum is given by
$\{\hbar[\omega_-(n_-+\frac{1}{2})+\omega_+(n_++\frac{1}{2})]~|~
n_\pm=0,1,2,\cdots\}$. This coincides with the spectrum of $\hat
H_\delta$, for $\hat H_\delta$ and $\hat H'_\delta$ are
isospectral.\footnote{In fact, in this case the similarity
transformation can be implemented using an anti-Hermitian generator
$\hat Q$. Therefore it is a unitary transformation that leaves the
spectrum invariant.}

Next, we note that according to (\ref{U=}),
    \be
    \det \mathbf{U}=\frac{4\gamma^2}{\gamma^2-\epsilon^2}.
    \label{det-U}
    \ee
Therefore, for a degenerate PU oscillator where $\gamma=0$, there is
no invertible matrix $\mathbf{U}$ that could implement the desired
linear canonical transformation. This means that in this case we
cannot decouple and diagonalize the Hamiltonian $\hat H_\delta$ by
performing a linear canonical transformation (a similarity
transformation with a quadratic generator $\hat Q$.) This suggests
that in this case $\hat H_\delta$ might even be
non-diagonalizable.\footnote{We are currently unable to establish if
this is actually true.}

\np

\ed
\begin{thebibliography}{99}

\bibitem{PU} A.~Pais and G.~E.~Uhlenbeck, Phsy.\ Rev.~{\bf 79}, 145
(1950).

\bibitem{stelle} K.\ S.\ Stelle, Phys.\ Rev.~D \textbf{16}, 953
(1977).

\bibitem{ma-da} P.~D.~Mannheim and A.~Davidson, Phys.\ Rev.~A {\bf
71}, 042110 (2005).

\bibitem{smilga-plb} A.~V.~Smilga, Phys.\ Lett.\ B {\bf 632}, 433
(2006).

\bibitem{be-ma} C.~M.~Bender and P.~D.~Mannheim, Phys.\ Rev.\ Lett.\
{\bf 100}, 110402 (2008).

\bibitem{jpa-2008b} A.~Mostafazadeh, J.\ Phys.\ A  \textbf{41}, 055304
(2008).

\bibitem{quasi} F.~G.~Scholtz, H.~B.~Geyer, and F.~J.~W.~Hahne,
Ann.\ Phys.\ (NY) {\bf 213} 74 (1992).

\bibitem{jpa-2003} A.~Mostafazadeh, J.~Phys.~A {\bf 36}, 7081
(2003).

\bibitem{cal} E.~Caliceti, F.\ Cannata and S.\ Graffi, J.\ Phys.\ A \textbf{39},
10019 (2006).

\bibitem{review}  A.~Mostafazadeh, preprint arXiv: 0810.5643, to appear in Int.\ J.~Geom.\
Meth.\ Mod.\ Phys.


\bibitem{bender-class} C.~M.~Bender and S.~Boettcher,
Phys.\ Rev.\ Lett.\ {\bf 80}, 5243 (1998); C.~M.~Bender,
S.~Boettcher, and P.~N.~Meisenger, J.~Math.\ Phys.\ {\bf 40}, 2201
(1999).

\bibitem{pla-2006} A.~Mostafazadeh, Phys.\ Lett.\ A \textbf{357} 177
(2006).

\bibitem{smilga} A.~V.~Smilga, J.\ Phys.\ A \textbf{41}, 244026 (2008).

\bibitem{CGS} E.~Caliceti, S.~Graffi, and J.~Sj\"ostrand, J.~Phys.\
A 40, 10155-10170 (2007).

\bibitem{BCF} B.~L.~Burrows, M.~Cohen, and T.~Feldmann, Int.\
J.~Quantum Chem.\ {\bf 92}, 345-354 (2003).

\end{thebibliography}
